\documentclass[aps,prd,twocolumn,showpacs,nofootinbib]{revtex4}
\usepackage{amsmath}
\usepackage{graphicx}
\usepackage{dcolumn}
\usepackage{bm}
\usepackage{amssymb}
\usepackage{latexsym}
\usepackage{bm}
\usepackage{float}

\newcommand{\no}{\nonumber}

\newcommand{\pa}{\partial}
\newcommand{\del}{\delta}
\newcommand{\eps}{\epsilon}

{\bf }

\newcommand{\half}{\frac{1}{2}}

\begin{document}


\title{
Chern-Simons  quantum mechanics and fractional angular momentum in atom system
}

\author{Yao-Yao Ma  ${}^{a}$}

\author{Qiu-Yue Zhang ${}^{a}$}

\author{Qing Wang ${}^{b}$}



\author{Jian Jing ${}^{a}$}
\email{jingjian@mail. buct. edu. cn}


\affiliation{${}^a$ Department of Physics and Electronic, School of
Science, Beijing University of Chemical Technology, Beijing 100029,
P. R. China, }

\affiliation{${}^b$ College of Physics and Technology, Xinjiang University, Urumqi 830046, P. R. China, }



\begin{abstract}

The model of a  planar atom which possesses a non-vanishing electric dipole moment interacting with magnetic fields in a specific setting is studied. Energy spectra of  this model and its reduced model,  which is the limit of cooling down the atom to the negligible kinetic  energy, are solved exactly. We show that 
energy spectra of the reduced model can not be obtained directly from the full ones by taking the same limit. In order to get the energy spectra of the reduced model from the full model, we must regularize  energy spectra of the full model properly when the limit of the negligible kinetic energy is taken. It is one of the characteristics of the Chern-Simons quantum mechanics.   Besides this, the canonical angular momentum of the reduced model will take fractional values although the full model can only take integers. It means  that it is possible to  realize the Chern-Simons quantum mechanics and fractional angular momentum simultaneously by this model.

\end{abstract}

{}

\maketitle

Chern-Simons quantum mechanics  is firstly studied in \cite{jackiw}. It describes a charged planar particle confined by a quadratic  potential in the background of a constant external magnetic field. It is found that the reduced model, which is the zero-mass  ($\mu \to 0$) limit of the full model, behaves non-trivially. Although the solutions to the classical equations of motion of the reduced model can be obtained directly from the full ones by taking the limit $\mu \to 0$ directly, energy spectra of this reduced model can not be obtained straightforwardly from the full model in the same manner.  Because  energy spectra of the full model will be divergent when the limit $\mu \to 0$ is taken. In order to match them,  
one must regularize spectra of the full theory artificially when this limit is taken. 
Besides it, the authors show that   eigenvalues of the canonical angular momentum take integer values  in the full theory (in the unit of $\hbar=1$). However, in the reduced model, it can only take positive half-odd numbers.

Ref \cite{baxter} studies  an atom with a non-vanishing electric dipole moment interacting with magnetic fields. It is found that  eigenvalues of the canonical angular momentum of the reduced model, which is obtained by cooling down the atom to the negligible kinetic energy, can only take  positive half-odd numbers. It is one of the characteristics of the  Chern-Simons quantum mechanics. It means that it is possible  to realize Chern-Simons quantum mechanics by a neutral particle which was realized by  a charged particle and a uniform magnetic field before.

The fractional angular momentum is an interesting issue in physics \cite{wilczek,liang}. It receives some renewed interests recently \cite{yzhang1,yzhang2,lr} . As is well-known, eigenvalues of the canonical angular momentum should be quantized in the unit of $\hbar/2$   in three-dimensional space due to the non-Abelian  rotation group. However, this conclusion is no longer true in $2+1$ dimensional space-time since the rotation group in two-dimensional space is an Abelian one which does not impose any constraints on  eigenvalues of the canonical angular momentum. Due to the  dynamical nature of the Chern-Simons gauge field and  the absence of the Maxwell term, one can realize the fractional angular momentum in $2+1$ dimensional space-time by coupling  charged particles or charged  fields with the Chern-Simons gauge field \cite{cs1, cs2, cs3, forte, banerjee1, banerjee2}. 

 Ref. \cite{zhang} provides an alternative  approach to realize the fractional angular momentum by using a cold ion.  The author  considers an trapped planar ion coupling to two different types of  magnetic potentials. One is the dynamical 
 the other  is the Aharonov-Bohm type. 
As expected,  eigenvalues of the canonical angular momentum of this model take integers. However, it is showed that eigenvalues of the canonical angular momentum of the reduced model, which is obtained by cooling down the kinetic energy of the ion to its lowest level, take fractional values. The fractional part  is proportional to the magnetic flux inside the solenoid.

In this paper, we propose a model to realize  Chern-Simons quantum mechanics and  fractional angular momentum simultaneously. Different from previous studies \cite{jackiw, zhang}  which realize  Chern-Simons quantum mechanics and fractional angular momentum by charged particles, we realize them simultaneously by a cold atom which   possesses a non-vanishing electric dipole moment. 

Our model is a planar atom with a non-vanishing electric dipole moment interacting with two magnetic fields. This atom is trapped by a harmonic potential and the  electric dipole moment is perpendicular to the plane. The  harmonic potential and the magnetic fields are arranged that  the motion of the atom  is  rotationally symmetric.
To be specific, the magnetic fields take the form
\begin{eqnarray}
\mathbf B^{(1)} &=& \frac{ \lambda}{2 \pi r} \mathbf e_r, \label{mfs1} \\
\mathbf B ^{(2)} &=& \frac{ \rho r}{2}  \mathbf e_r, \label{mfs2}
\end{eqnarray}
where 
$\lambda$ and $\rho$ are two parameters which characterize  intensities of these two magnetic field, 
$\mathbf e_r$ is the unit vector along the radial direction on the plane.

In three-dimensional space, the Hamiltonian which governs the dynamics of an atom with a non-vanishing electric dipole moment in the background of a magnetic field  $\mathbf B$ is 
\begin{equation}\label{ha1}
H = \frac{1}{2 \mu} (\mathbf p + \frac{d }{c^2} \mathbf n \times \mathbf B)^2 
\end{equation}
in which $\mu$ is the mass of the atom, $\mathbf p = -i \hbar \bm \nabla$ is the canonical momentum,  $d$ is the magnitude of the electric dipole moment and  $\mathbf n$ is the unit vector along the electric dipole moment. 
The Hamiltonian (\ref{ha1}) is the non-relativistic limit of a relativistic spin-half neutral particle with  a non-vanishing electric dipole moment interacting with  magnetic fields \cite{he1993}. In our model we apply two magnetic fields (\ref{mfs1}, \ref{mfs2}) simultaneously. Therefore the magnetic field in eq. (\ref{ha1}) is $\mathbf B = \mathbf B^{(1)} + \mathbf B^{(2)}$. 

One can derive the  He-Mckellar-Wilkens (HMW) effect \cite{he1993, wilkens} from the above Hamiltonian by only turning on the magnetic field $\mathbf B^{(1)}$. 
The HMW effect is firstly predicted by He and McKellar \cite{he1993} and later independently by Wilkens \cite{wilkens}.
This  effect predicts that a neutral particle with a non-vanishing electric dipole moment will accumulate a  topological phase if it moves around a line of magnetic charge with its electric dipole moment paralleling to the line. It is argued that the HMW effect is  electromagnetically dual to the AB effect. 
The magnetic field $\mathbf B^{(1)}$ plays an analogous role as the magnetic potentials produced by the long-thin magnetic flux-carried solenoid in the Aharonov-Bohm (AB) effect \cite{AB, eab, eab2}.   

The other  aspect of the Hamiltonian (\ref{ha1})  is that the energy eigenvalues of  Hamiltonian (\ref{ha1}) with only $\mathbf B^{(2)}$ turning on are analogous with Landau levels \cite{braz}. Thus, it means that the Landau levels can also be realized by a neutral particle with a non-vanishing  electric dipole moment. It may afford a possible  method to realize the quantum Hall effect \cite{qhe1, qhe2, qhe3} by neutral particles. 

It should be mentioned that the similarity between Landau levels and eigenvalues of neutral particles interacting with electromagnetic fields in various backgrounds had attracted much attention since the work of \cite{es}. In refs.  \cite{bakke1, bakke1a, bakke2, bakke3, bakke4, bakke5, bakke6, bakke7},
the authors solved energy spectra of neutral particles which  possesses  non-vanishing electric or magnetic dipole moments in the background of electromagnetic fields analytically in various configurations.

Since the motion of the atom is on a plane which is  perpendicular to the electric dipole moment,  we only concentrate on this plane. Furthermore, we trap the atom by a harmonic potential. So the model we consider is described by the Hamiltonian (the summation convention is applied) 
\begin{equation}
H = \frac{1}{2 \mu}(p_i - \frac{d}{c^2} \eps_{ij} B_j )^2 + \half K x_i ^2 
\label{ha2}
\end{equation}
where  $\half K x_i ^2$ is the harmonic potential with $K$ being a constant. The Lagrangian corresponding to this Hamiltonian  is
\begin{equation}
L = \half m \dot x_i ^2  + \frac{d}{c^2} \eps_{ij} \dot x_i B_j   - \half K x_i ^2.  \label{la1}
\end{equation}

The energy spectra of the Hamiltonian (\ref{ha2}) can be solved analytically. In doing so, we introduce  polar coordinates $(r, \ \theta) $ and substitute magnetic fields (\ref{mfs1}, \ref{mfs2}) into the Hamiltonian (\ref{ha2}). The eigenvalue  equation $H \psi =E \psi$ in the operator form becomes
\begin{eqnarray}
&&\big [-\frac{\hbar^2}{2\mu} (\frac{\pa ^2} {\pa r^2 } + \frac{1}{r} \frac{\pa }{\pa r} + \frac{1}{r^2} \frac{\pa ^2 }{\pa \theta^2}) + (\half \mu \omega ^2 + \frac{K}{2}) r^2   \no \\
&&  + \frac{d^2 \lambda^2}{8 \pi^2 \mu c^4 r^2} + \frac{d^2 \rho \lambda}{ 4 \pi \mu c^4} + (\frac{d \lambda}{2 \pi \mu c^2 r^2} + \omega) L_z  \big] \psi(r, \ \theta) \no \\ &&= E \psi (r, \ \theta), \label{eef1}
\end{eqnarray}
where 
$$\omega = \frac{d \rho}{2 \mu c^2}$$ 
is a parameter having the dimension of frequency,  $L_z = -i \hbar \frac{\pa}{\pa \theta}$ is the canonical angular momentum perpendicular to the plane. 

Since $L_z$ commutes Hamiltonian, i.e.,  $[L_z, \ H]=0$, we decompose the wavefunction as $\psi(r, \ \theta) = R(r) \Theta (\theta) = R(r) e^{im \theta}, \ m=0, \pm 1, \pm 2, \cdots$. Substitute it into eq. (\ref{eef1}), we get  equation the radial function $R(r)$  satisfied. It is 
\begin{eqnarray}
&&\big [\frac{d ^2} {d r^2 } + \frac{1}{r} \frac{d }{d r} - \frac{\alpha^2}{r^2}  
- (\frac{\mu^2 \omega^2}{\hbar^2} + \frac{\mu K}{\hbar^2}) r^2  \\ && + (\frac{2 m E}{\hbar^2} - \frac{d^2 \rho \lambda}{2 \pi c^4 \hbar^2} + \frac{2 m \mu \omega}{\hbar}) \big] R(r)=0. \label{radialeq}
\end{eqnarray}
where 
\begin{equation}
 \alpha^2 =\frac{d^2 \lambda ^2}{4 \pi ^2 c^4 \hbar^2} - \frac{m d \lambda}{\pi c^2 \hbar} + m^2
\end{equation}
is a dimensionless parameter. 
In order to simplify the radial equation, we introduce an auxiliary variable $\xi$ which relates $r^2$ by
\begin{equation}
\xi = \frac{ \sqrt{\mu^2 \omega^2 + \mu K}}{\hbar} r^2.
\end{equation}
In terms of $\xi$, we rewrite the radial equation (\ref{radialeq}) as
\begin{equation}
\xi \frac{d ^2R(\xi)}{d \xi^2} + \frac{d R(\xi)}{d \xi} + (\beta - \frac{\xi}{4} - \frac{\alpha^2}{4 \xi})R (\xi)=0 \label{radialeq2}
\end{equation}
where $\beta$ is a  parameter defined as
\begin{equation}
\beta = \frac{4 \mu E  \pi c^2 - d^2 \lambda \rho + 2 \pi  d \rho m \hbar c^2} {8 \pi \hbar  c^4 \sqrt{\mu K + \mu^2 \omega^2}}.
\end{equation}
After a careful analysis of asymptotic behaviors of equation (\ref{radialeq2}), one can find that the solution must take the form
\begin{equation}
R(\xi) = C e ^{- \half \xi} \xi ^{\half |\alpha|} \phi(\xi)
\end{equation}
where $C$ is the normalization constant to be determined and $\phi(\xi)$ is the solution  to the confluent hypergeometric equation
\begin{equation}
\xi \frac{d^2 \phi(\xi)}{d \xi ^2} + (|\alpha|+1 - \xi)  \frac{d \phi(\xi)}{d \xi} -(\frac{|\alpha|+1}{2} - \beta) \phi(\xi)=0.
\end{equation}
The solution to this confluent hypergeometric equation is 
\begin{equation}
\phi(\xi) = F \Big [ (\frac{|\alpha| +1}{2} -\beta  ), \ |\alpha|+ |, \ \xi \Big]. \label{hgf}
\end{equation}
Therefore, the solution to eq.  (\ref{eef1}) is 
\begin{eqnarray}
\psi(r, \ \theta) = &&\frac{1}{\lambda_0^{|\alpha|}} \big [\frac{\Gamma(n + |\alpha| +1)}{2 ^{|\alpha|+1} \pi n! \Gamma ^2 (|\alpha|+1) } \big] ^\half  e^{im \theta} r^{|\alpha|} \no \\
&&\times \exp (-\frac{r^2}{4 \lambda^2_0}) F(-n, |\alpha|+1, \frac{r^2}{2 \lambda_0})
\end{eqnarray}
where
\begin{equation}
\lambda_0 = \frac{\hbar}{\sqrt 2 (\mu ^2 \omega^2 + \mu K)^\frac{1}{4}}. \no
\end{equation}

The energy spectra are determined by  
setting the first parameter in the hypergeometric function (\ref{hgf}) to a negative integer number, i.e., 
\begin{equation}
 \frac{|\alpha| +1}{2} - \beta = -n, \quad n=0, 1, 2, \cdots.
\end{equation}
Thus, we get the energy spectra of Hamiltonian (\ref{ha2}) which are labeled by quantum numbers $n$ and $m$
\begin{eqnarray}
E_{n,m} = (2n+1 + |m|) \hbar \Omega - m \hbar \omega  - \frac{d \lambda }{2 \pi c^2} (\Omega - \omega) \label{spectra}
\end{eqnarray}
with $\Omega$ being
\begin{equation}
\Omega  = \sqrt {\omega^2 + \frac{K}{\mu}}.
\end{equation}

Now we study the reduced model which is the  negligibly small kinetic energy limit of model (\ref{ha2}). This reduced model can be realized in experiments by cooling down the atom to a very low speed \footnote{The speed of an atom can be cooled down to 1 m/s in the early of 1990s \cite{sst}. }.  We shall show that the model (\ref{ha2}) can realize Chern-Smions quantum mechanics and fractional angular momentum simultaneously when certain limit is taken.  

In the limit of the negligibly small kinetic energy,  Hamiltonian of the reduced model is deduced from (\ref{ha2}) by neglecting the kinetic energy term,  
\begin{equation}
H_r = \frac{K}{2} x_i ^2.  \label{reducedh}
\end{equation}
However, the energy spectra of the reduced model (\ref{reducedh}) can not be obtained directly from full one (\ref{spectra}) by taking the same limit. Mathematically, the limit of cooling down the kinetic energy to the negligibly small  amounts to take the limit of $\mu \to 0$. In this  limit, the frequency $\Omega$ becomes
\begin{equation}
\lim_{\mu \to 0} \Omega = \frac{d \rho}{2 \mu c^2} + \frac{c^2 K}{d\rho}.
\end{equation}
As a result,  the spectra (\ref{spectra}) become
\begin{eqnarray}
\lim_{\mu \to 0} E_{n, m} = && (2 n+ |m| -m +1) \frac{ \hbar d \rho}{2 \mu c^2} \no \\ && + (2n + |m| +1) \frac{\hbar c^2 K}{d \rho} - \frac{\lambda  K}{2 \pi \rho}. \label{rspectra}
\end{eqnarray}
Obviously, the spectra will be divergent when the kinetic energy of the atom is cooled down to the negligibly small. It means that energy spectra of the reduced model can not be obtained directly by setting the limit of $\mu \to 0$ from the full model although  there are no singularities in the  Lagrangian  (\ref{la1}) when the same limit is taken. It is one of the characteristics of the Chern-Simons quantum mechanics. 

In order to get physical results, we must analyze the spectra (\ref{rspectra}) carefully. The divergence of eigenvalues of states with $n>0$ and $n=0$, $m < 0$ can not be removed by a universal subtraction. However, eigenvalues  of states with $n=0, \ m > 0$ will be finite if one regularizes them by  a universal subtraction. This universal subtraction is $\frac{\hbar d \rho}{2 \mu c^2} + \half \frac{\hbar c^2 K}{d \rho} - \frac{\lambda K}{2 \pi \rho}$.  It means that  besides a infinite part $\frac{\hbar d \rho}{2 \mu c^2}$, one must remove a finite part $ \half \frac{\hbar c^2 K}{d \rho} - \frac{\lambda K}{2 \pi \rho}$ from (\ref{spectra}) during the reduction. Therefore,  the spectra of the reduced model should be  
\begin{equation}
E= \frac{\hbar K c^2}{d \rho} (|m| + \half). \label{regularized}
\end{equation}

For the sake of proving  that the regularization we made is reasonable, we come back to the  Lagrangian (\ref{la1}).

The limit of negligibly small kinetic energy demands us to  set the kinetic energy in (\ref{la1}) to zero. It leads to the reduced Lagrangian
\begin{equation}
L_r =\frac{d}{c^2}\eps_{ij} \dot x_i B_j - \frac{K}{2} x_i^2 . \label{reducedL}
\end{equation}

Introducing the canonical momenta with respect to $x_i$, we get
\begin{equation}
p_i = \frac{\pa L_r}{\pa \dot x_i} = \frac{d}{c^2} \eps_{ij}  B_j. \label{cmr}
\end{equation}
Evidently, the introduction of canonical momenta leads to primary constraints (the symbol `$\approx$' is `weak' equivalence which means equivalence only on the constraint surface)
\begin{equation}
\phi_i ^{(0)} = p_i - \frac{d}{c^2} \eps_{ij} B_j \approx 0 \label{pcs}
\end{equation}
in the terminology of Dirac \cite{dirac}. In fact, these constraints can also be gotten from Hamiltonian (\ref{ha1}) since in the limit of $\mu \to 0$ one must set $p_i -\frac{d}{c^2} \eps_{ij} B_j =0 $ in order to keep the Hamiltonian finite. 

The Poisson brackets among primary constraints $\phi^{(0)}_i$ are
\begin{equation}
\{ \phi_i ^{(0)}, \ \phi_j ^{(0)} \}= -\frac{ d \rho}{c^2} \eps_{ij}.
\end{equation}
Since $ \{ \phi_i ^{(0)}, \ \phi_j ^{(0)} \} \neq 0$, they belong to the second class and there are no secondary constraints. Because of the second class nature, they can be applied to eliminate the redundant degrees of freedom in the reduced model (\ref{reducedL}).  

Eigenvalues of the reduced Hamiltonian (\ref{reducedh}) can be obtained once the commutation relations between $x_i$ are determined. The classical version of  commutators among $x_i$, i.e.,  Dirac brackets among variables $x_i$ are defined by \cite{dirac}
\begin{equation}
\{x_i, \ x_j \}_D = \{x_i, \ x_j \} - \{x_i, \ \phi_m^{(0)} \} \{\phi_m ^{(0)}, \ \phi_n ^{(0)} \}^{-1} \{\phi^{(0)}_n, \ x_j \}.
\end{equation}
After some algebraic calculations, we arrive at
\begin{equation}
[x_i, \ x_j] = i \hbar \{x_i, \ x_j \} _D =  \frac{i\hbar c^2}{ d \rho } \eps_{ij}.  \label{comm2}
\end{equation}
Thus, the reduced Hamiltonian (\ref{reducedh}) is equivalent to a one-dimensional harmonics oscillator. Its eigenvalues can be read directly from the Hamiltonian (\ref{reducedh}) and commutation relation (\ref{comm2}) as
\begin{equation}
E_{n} =  \frac{\hbar c^2 K}{ d \rho } (n + \half), \ \ \ n=0,1,2, \cdots. \label{esrm}
\end{equation}
It coincides with  energy spectra  (\ref{regularized})  after the regularization. Thus, we show that the Chern-Simons quantum mechanics can be realized by the model (\ref{ha2}).

The other  feature of the model (\ref{ha2}) we shall study is its rotation property. We shall show that although the eigenvalues of the canonical  angular momentum of the model (\ref{ha2}) take conventional values, the canonical angular momentum of its reduced model (\ref{la1}) can take fractional values. 

The canonical angular momentum of the model (\ref{ha2}) is 
\begin{equation}
J = \eps_{ij} x_i p_j . \label{camf}
\end{equation}
Eigenvalues of canonical angular momentum are obvious integers since it can also be written as $J= -i \hbar \pa/\pa \theta$. By requiring the single-valuedness of wavefunctions, we conclude that $J_n = n \hbar, \ n=0, \pm 1, \pm2, \cdots$.   However, in the reduced model (\ref{reducedh}) or (\ref{reducedL}), because of intrinsic  constraints (\ref{pcs}),  variables $x_i$ and $p_i$ are not independent. Substituting the primary constraints (\ref{pcs}) into (\ref{camf}), we get the canonical angular momentum of the reduced model
\begin{eqnarray}
J_r = \eps_{ij} x_i p_j 
 = -\frac{ d}{2 c^2} (\rho x^2 _i+ \frac{\lambda}{\pi}). \label{rcam}
\end{eqnarray}
Using the commutation relations among $x_i$ (\ref{comm2}), we can verify straightforwardly that the canonical angular momentum (\ref{rcam}) not only is the generator of the rotation transformation
\begin{equation}
[J_r, \ x_i] = i \hbar \eps_{ij} x_j.
\end{equation}
but also is conserved since $[J_r, \ H_r]=0$. In fact, the canonical angular momentum (\ref{rcam}) is the Noether charge of the rotation transformation $\del x_i \sim \eps_{ij} x_j \del \theta$ with $\delta \theta$ being an infinitely small angle. 

Up to a constant, the canonical angular momentum of the reduced model is analogous to a one-dimensional harmonic oscillator. With the help of commutators (\ref{comm2}), we can get  eigenvalues of $J_r$ easily. They are
\begin{equation}
J_{rn} = - (n+ \half ) \hbar - \frac{ d \lambda}{2 \pi c^2}, \ \ n=0,1,2, \cdots. \label{fcam}
\end{equation}

The result (\ref{fcam})  shows that eigenvalues of the canonical angular momentum can take fractional values. Apart from a minus sign, the fractional part is proportional to  $\lambda$, which is a  tunable classical parameter. 

To summarize, we propose a model to realize the Chern-Simons quantum mechanics and fractional angular momentum simultaneously. Different from previous work in which the Chern-Simons quantum mechanics and the fractional angular momentum are mostly realized by charged particles, we realize them by using a trapped  atom which possesses a non-vanishing electric dipole moment and two magnetic fields. This scheme  is dual to the one proposed in ref. \cite{jzd}, in which it was suggested  to use a neutral particle possessing a magnetic dipole moment and two electric fields to realize fractional angular momentum. The duality between the present one and the one proposed by ref. \cite{jzd} coincides with the electromagnetic duality in Maxwell equations \cite{jackson}.

\section*{Acknowledgements}

This work is supported by NSFC with Grant No. 11465006.

\end{document}